\documentclass{ifacconf}

\usepackage[utf8]{inputenc}

\usepackage{graphicx}      
\usepackage{natbib}        % required for bibliography
 \usepackage{amsmath} 
\usepackage{amsfonts}
\usepackage{amssymb}
\usepackage{xcolor}
\usepackage{float}  
\usepackage{upgreek}
\usepackage{mathtools}
\usepackage{float}
\usepackage{tikz}
    \usetikzlibrary{arrows}
\usepackage{soul}
\usepackage{IEEEtrantools}
\usepackage{enumerate}

\def\begequarr{\begin{eqnarray}}
\def\endequarr{\end{eqnarray}}
\def\begequarrs{\begin{eqnarray*}}
	\def\endequarrs{\end{eqnarray*}}
\def\begarr{\begin{array}}
	\def\endarr{\end{array}}
\def\begequ{\begin{equation}}
\def\endequ{\end{equation}}
\def\lab{\label}
\def\begdes{\begin{description}}
	\def\enddes{\end{description}}
\def\begenu{\begin{enumerate}}
	\def\begite{\begin{itemize}}
		\def\endite{\end{itemize}}
	\def\endenu{\end{enumerate}}

\def\lef[{\left[\begin{array}}
	\def\rig]{\end{array}\right]}
\def\qed{\hfill$\Box \Box \Box$}
\def\begcen{\begin{center}}
	\def\endcen{\end{center}}
\def\begrem{\begin{remark}\rm}
	\def\endrem{\end{remark}}
\def\begdef{\begin{definition}}
	\def\enddef{\end{definition}}
\def\begpro{\begin{proposition}}
	\def\endpro{\end{proposition}}
\def\begfac{\begin{fact}}
	\def\endfac{\end{fact}}
\def\begass{\begin{assumption}}
	\def\endass{\end{assumption}}

\def\begmat#1{\begin{bmatrix}#1\end{bmatrix}}
\def\begali#1{\begin{align}{#1}\end{align}}
\def\begalis#1{\begin{align*}{#1}\end{align*}}
\def\cali{{\cal I}}

\def\call{{\cal L}}

\def\caly{{\cal Y}}

\def\liminf{\lim_{t \to \infty}}
\def\callinf{{\cal L}_\infty}
\def\L2e{{\cal L}_{2e}}

\def\rea{\mathbb{R}}

\def\adj{\mbox{adj}}
\def\col{\mbox{col}}

\def\et{\varepsilon_t}

\usepackage[prependcaption,colorinlistoftodos]{todonotes}

\newtheorem{definition}{Definition}

\def\et{{\boldsymbol\eta}}

\begin{document}
    \begin{frontmatter}
        \title{A Globally Convergent State Observer for Multimachine Power Systems with Lossy Lines
        } 
        \thanks[footnoteinfo]{This work was partially supported by the Government of Russian
Federation (Grant 08-08), by the Ministry of Science and Higher Education of Russian Federation, passport of goszadanie no. 8.8885.2017/8.9, NSFC (61473183, U1509211).}

        \author[First]{Alexey Bobtsov} 
        \author[Second,First]{Romeo Ortega} 
        \author[First]{Nikolay Nikolaev}
        \author[Fourth]{Johannes Schiffer}
       
        \address[First]{Department of Control
Systems and Robotics, ITMO University, Kronverkskiy av. 49, Saint
Petersburg, 197101, Russia\\ (e-mail: bobtsov@mail.ru, nikona@yandex.ru)}
        \address[Second]{LSS-Supelec, 3, Rue Joliot-Curie, 91192 Gif-sur-Yvette, France\\ (e-mail: ortega@lss.supelec.fr)}
        \address[Fourth]{Fachgebiet Regelungssysteme und Netzleitechnik, Brandenburgische Technische Universität Cottbus - Senftenberg\\ (e-mail: schiffer@b-tu.de)}

         \begin{abstract}                % Abstract of not more than 250 words.
        We present the first solution to the problem of estimation of the state of multimachine power systems with lossy transmission lines. We consider the classical three-dimensional ``flux-decay" model of the power system and assume that the active and reactive power as well as the rotor angle and excitation voltage at each generator is available for measurement---a scenario that is feasible with current technology.  The design of the observer relies on two recent developments proposed by the authors: a parameter estimation based approach to the problem of state estimation and the use of the dynamic regressor extension and mixing technique to estimate these parameters. Thanks to the combination of these techniques it is possible to overcome the problem of lack of persistent excitation that stymies the application of standard observer designs. Simulation results illustrate the performance of the proposed observer.

        \end{abstract}


    \end{frontmatter}
%===============================================================================
\section{Introduction} 
\lab{sec1}
%%%%%%%%%%%%%%%%%%5
%
Power systems are experiencing major changes and challenges, such as an increasing amount of power-electronics-interfaced equipment, growing transit power flows and fluctuating (renewable) generation, see \citep{ winter15}. Therefore power systems are operated under more and more stressed conditions and, thus, closer to their stability limits as ever before, see \citep{ winter15}. In addition, as detailed in \citep{ milano18}, their dynamics become faster, more uncertain and also more volatile. Hence, fast and accurate monitoring of the system states is crucial in order to ensure a stable and reliable system operation, see \citep{ zhao19}.
This, however, implies that the conventional monitoring approaches based on steady-state assumptions are no longer appropriate and instead novel {\em dynamic state estimation (DSE)} tools have to be developed, see \citep{ zhao19,singh2018dynamic}.

By recognizing this need, DSE has become a very active research area in the past years, see \citep{ zhao19,singh2018dynamic}. The interest in DSE has been further accelerated by the growing deployment of phasor measurement units (PMUs), see \citep{ terzija10}.
As of today, the prevalent algorithms in the literature for DSE are Kalman Filter techniques, including Extended Kalman Filters, see \citep{ ghahremani11,paul18} and Unscented Kalman Filters, see \citep{ valverde11,wang11,anagnostou17}. The main reasoning behind using these techniques is that they are in principle applicable to nonlinear systems, see \citep{ julier97,wan00}, such as power system models. However, a major drawback of the aforementioned results is that the estimator convergence is only shown empirically, {\em e.g.}, via simulations, but no rigorous convergence guarantees are provided.

An important application of DSE is the design and practical implementation of excitation controllers for transient stability improvement, e.g., via power system stabilizers, see \citep{ zhao19}. To the best of our knowledge, all existing solutions to this problem with rigorous stability claims rely on the availability of the full state of the system. The assumption that all generating units dispose of full state measurement severely stymies the practical applicability of these controllers. Besides its potential application in excitation controllers, the present result may be utilized by the wide-area control system to control the transient and oscillatory dynamics of system voltage and frequency \citep{ singh2018dynamic}. See the recent books by \citep{ sauer17,singh2018dynamic} for a detailed treatment on this new technological developments.

Recently, power system state observers have been proposed in \citep{ rinaldi17,rinaldi18,rinaldi19} based on sliding mode techniques and their convergence has been established under certain assumptions. Yet, all these results neglect the voltage dynamics and assume purely inductive transmission lines. In \citep{ anagnostou18} an observer-based anomaly scheme has been derived using a detailed linearized power system model in combination with a linear time-varying observer.

The abovementioned facts and developments motivate the observer design conducted in the present paper. For its derivation, we consider a large-scale power system consisting of $n$ synchronous generators represented by the standard three-dimensional "flux-decay" model, see \citep{ kundur94,sauer17}, and interconnected through a transmission network, which we assume to be lossy. That is, we explicitely take into account the presence of transfer conductances. It is well known that the dynamics of power systems with lossy transmission lines is considerably more complicated than the lossless one---see \citep{ ANDFOU,ORTetal} for a discussion on this issue. 

The main contribution of the present work is the proof that using the {\em measurements} of active and reactive power as well as rotor angle and excitation voltage at each generator---a reasonable assumption given the latest technology, see \citep{ yang07}---it is possible to recover {\em the full state} of a multimachine power system, even in the presence of lossy lines. An important observation that is made in the paper is that, under the assumed measurement scenario, the problem of estimation of the generator voltages can be recast as estimation of the state of a {\em linear time-varying} (LTV) system, to which classical state estimation techniques, {\em e.g.}, Kalman-Bucy filters \citep{ AND}, can be applied. Unfortunately, we prove in the paper that the associated LTV system is not uniformly completely observable (UCO), a necessary condition to ensure convergence of the estimated states, see  \citep{ rueda2019gramian} for a recent discussion of this issue and \citep{ zhao19} for a general discussion on the challenges of observability analysis in DSE. 

To overcome this problem we invoke two recent results reported by the authors. 
\begenu
\item Following the parameter estimation based observer (PEBO) design proposed in \citep{ ORTetalscl}, we reformulate the problem of state observation into one of {\em parameter} estimation. 
\item The unknown parameters are reconstructed using the dynamic regressor extension and mixing (DREM) procedure \citep{ ARAetal}, which is a novel, powerful, parameter estimation technique particularly suitable for PEBO design problem.
\endenu 
See \citep{ORTetaljpc,PYRetal} for two recent applications of these combined techniques. Two outstanding features of DREM estimators that are exploited in the paper are that, on one hand, no appeal is made to persistent excitation (PE) arguments to prove its convergence\footnote{We recall that PE of the classical linear regression model is equivalent to non UCO of the associated LTV system \citep{SASBOD}.}. Instead, a condition of non-square integrability of a scalar signal is imposed in DREM, see \citep{ARAetal} for further details. On the other hand, as shown in \citep{GERetal}, a simple modification to DREM allows to achieve the very important feature of {\em finite-time convergence} (FTC) under the weakest sufficient excitation assumption \citep{KRERIE}.  The convergence conditions mentioned above are, of course, necessary since as is well-known some excitation assumptions are needed to ensure the success of state observers \citep{BES,BER} and parameter estimators \citep{LJU,SASBOD}.

The remainder of the paper is organized as follows. The mathematical model of the power system and the problem formulation is given in Section \ref{sec2}. The proposed observers and their stability proof are given in  Section \ref{sec3}. Simulation results with a two machine example, which illustrate the {performance} of the proposed observer, are presented in Section \ref{sec4}. The paper is wrapped-up with concluding remarks in  Section \ref{sec5}.

%%%%%%%%%%%%%%%%%%%%
\section{System Model and State Observer Problem Formulation} 
\label{sec2}
%%%%%%%%%%%%%%%%%%%%%%%%%%%%%%%%%%%%%%%%%%%%%%%%%%%%%%%%%%%%%%%%%%%%%%%%%%%%%%%%%
%
As standard in centralized DSE, see \citep{ zhao19}, we consider a Kron-reduced power system consisting of $n>1$ interconnected machines and with the dynamics of the $i$--th generator described by the classical third order model\footnote{To simplify the notation, whenever clear from the context, the qualifier ``$i\in \bar n$" will be omitted in the sequel.}  \citep{ ANDFOU,kundur94,sauer17}
\begin{footnotesize}
\begin{equation}
\begin{split}
\dot{\delta}_i&=\omega_i, \\
M_i\dot{\omega}_i&=-D_{mi}\omega_i+\omega_0(P_{mi}-P_{ei}),\\
\tau_i\dot{E}_i&=-E_i-(x_{di}-x'_{di})I_{di}+E_{fi}+\nu_i,\;\;\;i\in \bar n:=\{1,...,n\},
\label{sys}
\end{split}
\end{equation} 
\end{footnotesize} 
where the state variables  are the rotor angle $\delta_i\in [0,2\pi)$, $\mbox{rad}$, the speed deviation $\omega_i\in\mathbb{R}$ in $\mbox{rad/sec}$ and the generator
quadrature internal voltage $E_i\in\mathbb{R}_+$, $I_{di}$ is the $d$ axis current, $P_{ei}$ is the electromagnetic power, the voltages $E_{fi}$ and
$\nu_i$ are the constant voltage component applied to the field winding, and the control voltage input, respectively. Furthermore, $D_{mi}$, $M_i$, $P_{mi}$,  $\tau_i$, $\omega_0$, $x_{di}$ and $x'_{di}$ are positive parameters. 

The active $P_{ei}$ and reactive $Q_{ei}$ power are defined as
\begin{equation}
\begin{split}
P_{ei}&=E_i I_{qi}\\
Q_{ei}&=E_i I_{di},
\label{peqe}
\end{split}
\end{equation}
where $I_{qi}$ is the $q$ axis current.

These currents establish the connections between the machines and are given by 
\begin{equation} 
\begin{split}
I_{qi}&=G_{mii}E_{i}+ \sum_{j=1,j\neq i}^{n} E_{j} Y_{ij} \sin(\delta_{ij}+\alpha_{ij})\\
I_{di}&=-B_{mii}E_{i}-\sum_{j=1,j\neq i}^{n} E_{j} Y_{ij} \cos(\delta_{ij}+\alpha_{ij}),
\label{idiq}
\end{split}
\end{equation}
where we defined $\delta_{ij}:=\delta_i-\delta_j$ and the constants $Y_{ij}=Y_{ji}$ and $\alpha_{ij}=\alpha_{ji}$ are the admittance magnitude and admittance angle of the power line connecting nodes $i$ and $j$, respectively. Furthermore, $G_{mii}$ is the shunt conductance and $B_{mii}$ the shunt susceptance at node $i$. Finally, combining (\ref{sys}), \eqref{peqe} and (\ref{idiq}) results in the well-known compact form \citep{ ORTetal,LANetal}
\begin{footnotesize}
\begin{equation} 
\begin{split}
\dot{\delta}_i &=  \omega_i\\
\dot{\omega}_i &=  -D_i\omega_i+P_i-d_i\Big[G_{mii}E_i^2- E_i\sum_{j=1,j\neq i}^n E_j Y_{ij}\sin(\delta_{ij}+\alpha_{ij})\Big]\\
\dot{E}_i &=  -a_iE_i+b_i\sum_{j=1,j\neq i}^n E_j Y_{ij}\cos(\delta_{ij}+\alpha_{ij})+ u_i,
\label{syscom}
\end{split}
\end{equation}
\end{footnotesize}
where we have defined the signals
$$
u_i:= \frac{1}{\tau_i}(E_{f_i}+\nu_i)
$$
and the positive constants 
\begalis{
	D_i&:=\frac{D_{mi}}{M_i},\; P_i:=d_i P_{mi},\; d_i:=\frac{\omega_0}{M_i}\\
	a_i&:=\frac{1}{\tau_i}[1-(x_{di}-x'_{di})B_{mii}],\;b_i:=\frac{1}{\tau_i}(x_{di}-x'_{di}).
}

To formulate the observer problem we consider that all parameters are known, and make the following assumption on the available {\em measurements}.

\begin{assum} 
	\label{ass1}
	The signals $\delta_i$, {$u_i$,} $P_{ei}$ and $Q_{ei}$ of all generating units are measurable.
\end{assum}

As usual in observer problems \citep{ BES,BER}  we assume that $u_i$ is bounded and such that all state trajectories are {\em bounded.}

\begdes
\item {\bf Problem Formulation:} Consider the multimachine power system (\ref{syscom}), verifying Assumption \ref{ass1}. Design an observer\footnote{For a set of scalar quantities $x_i$ we define the vector $x:=\col(x_1,x_2,\dots,x_n)$.}  
\begalis{
	\dot \chi &= F(\chi,\delta,P_{e},Q_{e})\\
	\begmat{\hat E \\ \hat \omega}&:=H(\chi,\delta,P_{e},Q_{e},\hat E, \hat \omega)
}
such that
\begequ
\lab{estcon}
\liminf \begmat{\tilde E(t) \\ \tilde \omega(t)}=0,
\endequ
where we, generically, define the estimation errors $\tilde{(\cdot)}:=\hat{(\cdot)}-(\cdot)$.
\enddes

Although the centralized measurement of the rotor angles $\delta_i$ is unlikely in applications, the observer problem without this assumption is a daunting task. Even if this transfer of information is possible, another practical consideration that is neglected in the problem formulation is the difficulty of synchronizing these measurements \citep{ singh2018dynamic,sauer17}.

%%%%%%%%%%%%5
\section{Main Result}
\lab{sec3}
%%%%%%%%%%%%%%
%
Before presenting the proposed observer we first give in this section an LTV systems perspective of the voltage dynamics and show its fundamental lack of observability limitation. To overcome the latter, we then adopt a PEBO perspective for the state observation problem and apply the DREM procedure to estimate the unknown parameters.

\subsection{An LTV description of the voltage dynamics}
\lab{subsec31}
%%%%%%%%%%%%%%
%
To simplify the notation we find convenient to introduce the following notation
\begin{footnotesize}
\begali{
\nonumber
\Delta_{ij}(t) & :=\delta_i(t)-\delta_j(t)+\alpha_{ij}\\
A(t) &:=
\begmat{
-a_1& b_1 Y_{12}\cos\Delta_{12}(t)& \dots & b_1 Y_{1n}\cos\Delta_{1n}(t) \\
b_2 Y_{21}\cos\Delta_{21}(t)& -a_2& \dots & b_2 Y_{2n}\cos\Delta_{2n}(t)\\
\vdots& \vdots& \vdots& \vdots \\
b_n Y_{n1}\cos\Delta_{n1}(t)& b_n Y_{n2}\cos\Delta_{n2}(t)&  \dots & -a_n
}.
\lab{a}
} 
\end{footnotesize}

Given this notation we observe that we can write the voltage equations of \eqref{syscom} as an {\em LTV system}
\begequ
\lab{dote}
\dot E = A(t)E+u.
\endequ 
We underscore that, under the standing assumptions, $A(t)$ is {\em measurable} and  obviously, bounded. 

The following simple lemma is instrumental for our further developments.

\begin{lem}
\lab{lem1}
There exists a {\em measurable} matrix $C(t):=C(P_e(t),Q_e(t),\delta(t))\in \rea^{n \times n}$ such that
\begequ
\lab{ce}
C(t)E=0
\endequ
\end{lem} 

\begin{pf}
From \eqref{peqe} we have that
$$
P_eI_d - Q_eI_q=0.
$$
Clearly, the equations \eqref{idiq}---which are linearly dependent on $E$---may be written in the compact form
\begequ
\lab{idiqcom}
I_q=S(\delta)E,\;I_d=T(\delta)E,
\endequ 
for some suitably defined $n \times n$ matrices $S(\delta)$ and $\;D(\delta)$. The proof is completed replacing \eqref{idiqcom} in the identity above and defining
\begequ
\lab{cexp}
C(P_e,Q_e,\delta):=\begmat{P_{e1}T_1^\top (\delta)-Q_{e1}S_1^\top (\delta) \\ \vdots \\ P_{en}T_n^\top (\delta)-Q_{en}S_n^\top (\delta)},
\endequ
where $T_i^\top (\delta),\;S_i^\top (\delta)$ are the rows of the matrices $D(\delta)$ and $S(\delta)$, respectively.
\qed
\end{pf}

One might be tempted to view the system \eqref{dote} and \eqref{ce} as an LTV system with output $y_E:=C(t)E$ that---in view of \eqref{ce}---turns out to be {\em equal to zero}, and design a classical Kalman-Bucy filter for it \citep{ AND}. That is, to propose the Kalman-Bucy observer 
\begin{footnotesize}
\begalis{
\dot{\hat  E} & = [A(t)-H(t)C^\top (t)C(t)]\hat E +u\\
\dot H &= HA^\top (t)+A(t)H-HC^\top (t)C(t)H+S,\;H(0)>0,\;S>0,
}
\end{footnotesize}
which ensures $\liminf \tilde E(t)=0$ (exponentially), provided the pair $(A(t),C(t))$ is UCO---see \citep{ AND} and the derivations in Section 5 of \citep{ rueda2019gramian}.\footnote{Actually, it is shown in  \citep{ AND} that the UCO condition ensures the Riccati equation above has a bounded, invertible solution $H$, which is necessary for the analysis of the associated observer error dynamics.} Alas, this system turns out to be non UCO. Indeed, we recall the following \citep{ SASBOD}

\begin{defn}
\lab{def1}
An LTV system $\dot x=A(t)x,\;y=C(t)x$, with bounded matrices $A(t),C(t)$ is {\em uniformly completely observable} (UCO) if there exists positive constants $c_1, c_2$ and $T$ such that
\begin{footnotesize}
$$
c_1 \cali_n \geq \int_{t_0}^{t_0+T}\Phi^\top (\tau,t_0)C^\top (\tau)C(\tau)\Phi(\tau,t_0)d\tau \geq c_2 \cali_n,\;\forall t_0 \geq 0,
$$ 
\end{footnotesize}
where $\Phi(t,t_0)$ is the system's state transition matrix.
\end{defn}

Now, we recall that the state trajectories of the system $\dot x=A(t)x$ with initial condition $x(t_0) \in \rea^n$ satisfy  \citep{ DEM,RUG}
\begequ
\lab{ephi}
x(t)=\Phi(t,t_0)x(t_0),\;\forall t \geq t_0.
\endequ
In our scenario $C(t)x(t)=0$, hence, it is clear that the lower bound on the inequality of Definition \ref{def1} cannot be satisfied.
\subsection{An PEBO approach to the estimation of the voltage}
\lab{subsec32}
%%%%%%%%%%%%%%
%

The lemma below illustrates how, applying the PEBO approach proposed in \citep{ ORTetalscl}, we can generate a linear regression equation (LRE) where estimation of the unknown parameters leads to estimation of the unknown state $E$.

\begin{lem}
\lab{lem2}
Consider the dynamic equations of the voltage  \eqref{a}-\eqref{ce} and the dynamic extension 
\begali{
\nonumber
\dot \xi_E & = A(t) \xi_{E} + u\\
\dot \Phi &= A(t) \Phi,\;\Phi(0)=\cali_n.
\lab{peboe}}
There exists a {\em constant} vector $\theta \in \rea^n$, and {\em measurable} signals $y \in \rea^n$ and $\psi \in \rea^{n \times n}$  such that 
\begali{
\lab{exipsi}
E &=\xi_E - \Phi \theta\\
\lab{lre}
y&=\psi \theta.
}  
\end{lem}

\begin{pf}
Define the signal $e:=\xi_E-E$. From \eqref{dote} and \eqref{peboe} we see that $e$ satisfies
$$
\dot e= A(t)e.
$$
From the equation above and the properties of the state transition matrix $\Phi$ of $A(t)$ given in \eqref{ephi}, we conclude  that there exists a {constant} vector\footnote{Clearly, we have that $\theta:=e(0)$.} $\theta \in \rea^n$ such that  $e=\Phi\theta$ and, consequently, \eqref{exipsi} holds. 

To establish \eqref{lre} we multiply \eqref{exipsi} by $C(t)$ and, recalling \eqref{ce}, obtain the equation
$$
C(t)\xi_E=C(t)\Phi \theta.
$$
The proof is completed defining 
\begali{
\nonumber
y&:=C(t)\xi_E\\
\lab{ypsi}
\psi&:=C(t)\Phi.
}
\qed
\end{pf}

\subsection{Generation via DREM of $n$ scalar LRE for the unknown parameters $\theta_i$}
\lab{subsec33}
%%%%%%%%%%%%%%
%
In view of Lemma \ref{lem2} the only remaining step to construct the observer for $E$ is to, proceeding from the linear regression \eqref{lre},  propose an estimator for $\theta$. Here, again, we are confronted with the lack of UCO mentioned in Subsection \ref{subsec31}. Indeed, the UCO condition of Definition \ref{def1} is equivalent to the PE of the regressor $\Psi$. Hence, standard gradient or least squares estimators will not ensure (exponential) convergence \citep{ SASBOD}. To bypass this difficulty we invoke the DREM parameter estimation procedure proposed in \citep{ ARAetal}.

Although the construction of DREM allows for the use of general, LTV, $\callinf$-stable operators, for the sake of simplicity we consider here the use of simple LTI filters. Towards this end,  we propose a stable transfer matrix $F(s) \in \rea^{n \times n}(s)$, and define the signals
\begali{
\nonumber
{\bf Y}&:=F(p)y \in \rea^n\\
\nonumber
\Psi&:=F(p) \psi \in \rea^{n \times n}\\
\nonumber
\caly &:=\adj\{\Psi\}{\bf Y} \in \rea^n\\
\Delta&:=\det\{\Psi\} \in \rea,
\lab{drem}
}
where $p:={d \over dt}$ and $\adj\{\cdot\}$ is the adjugate operator.

\begin{lem}
\lab{lem3}
Consider the LRE  \eqref{lre} and the signals \eqref{drem}. Then, the following {\em scalar} LREs hold
\begali{
\caly_i & = \Delta \theta_i.
\lab{scalre}
}  
\end{lem}

\begin{pf}
Applying the filter $F(p)$ to the LRE  \eqref{lre} we obtain 
$$
{\bf Y}=\Psi \theta +\et,
$$
where $\et$ is an exponentially decaying term that, following the standard procedure, is neglected in the sequel. The proof is completed by multiplying the equation above by $\adj\{\Psi\}$ and recalling that for any, {\em possibly singular}, $n \times n$ matrix $B$ we have $\adj\{B\}B=\det\{B\}\cali_n$. 
\qed
\end{pf}

\subsection{Proposed state observers}
\lab{subsec34}
%%%%%%%%%%%%%%
%
We are now in position to present our main result: two globally convergent observers for the state of the multimachine power system (\ref{syscom}). The first one ensures {\em asymptotic} convergence while the latter guarantees FTC. In both cases, the required excitation conditions are rather weak. 

\begin{prop}
\lab{pro1}
Consider the multimachine power system (\ref{syscom}), verifying Assumption \ref{ass1}. Fix an $n \times n$ stable transfer matrix $F(s)$ and $2n$ positive numbers $\gamma_i$ and  $k_{\omega i}$. The voltage state observer defined by  \eqref{a}, \eqref{cexp}, \eqref{peboe}, \eqref{ypsi}, \eqref{drem}, together with
\begali{
\nonumber
\dot {\hat \theta}_i &=-\gamma_i \Delta (\Delta \hat{\theta}_i-\caly_i)\\
\hat E &=\xi_E-\Phi \hat \theta,
\lab{volstaest}
} 
and the speed observer given by
\begali{
\nonumber
\dot\xi_{\omega_i} &=   -D_i\hat \omega_i+P_i-d_iP_{ei} - k_{\omega i} \hat{\omega}_i\\
\hat \omega_i &=\xi_{\omega_i}+ k_{\omega i} \delta_i,
\lab{omestaobs}
} 
ensure \eqref{estcon}, with all signals bounded, provided $\Delta \notin \call_2$. 
\end{prop}

\begin{pf}
Replacing \eqref{lre} in the parameter estimator equation  yields
\begequ
\lab{tetaSol}
\dot {\tilde \theta}_i =-\gamma_i \Delta^2 \tilde \theta_i.
\endequ
Given the standing assumption on $\Delta$ we have that $\tilde \theta(t) \to 0$. Hence, invoking \eqref{exipsi} and \eqref{volstaest}, this implies that  $\tilde E(t) \to 0$. 

To prove the convergence of the speed estimator notice that, using \eqref{peqe} and \eqref{idiq}, the rotor speed dynamics \eqref{syscom} may be written as
$$
\dot{\omega}_i =  -D_i\omega_i+P_i-d_i P_{ei}.
$$
Then, compute from \eqref{syscom} and \eqref{omestaobs} the error dynamics 
$$
\dot {\tilde \omega}_i =  -(D_i+k_{\omega i})\tilde \omega_i.
$$
This completes the proof of the proposition.
\qed
\end{pf}

As shown in \citep{ GERetal}, a simple modification to DREM allows to achieve FTC for the voltage observer, under the weakest {\em sufficient excitation} assumption \citep{ KRERIE}, which is articulated in the assumption below. For ease of presentation, and without loss of generality, we assume that all the adaptation gains $\gamma_i$ in the parameter estimator \eqref{volstaest} are equal to $\gamma >0$.

\begin{assum}
\lab{ass2}
Fix a small constant $\mu \in (0,1)$. There exists a time $t_c>0$ such that
\begequ
\lab{conint}
\int_0^{t_c} \Delta^2(\tau) d\tau \geq - {1 \over \gamma} \ln(1-\mu).
\endequ 
\vspace{0.01cm}
\end{assum}

\begin{prop}
\lab{pro2}
Consider the multimachine power system (\ref{syscom}), verifying Assumption \ref{ass1}. Fix an $n \times n$ stable transfer matrix $F(s)$, a positive numbers $\gamma$ and $\mu \in (0,1)$. The voltage state observer defined by  \eqref{a}, \eqref{cexp}, \eqref{peboe}, \eqref{ypsi}, \eqref{drem}, together with
\begali{
\nonumber
\dot {\hat \theta}_i &=-\gamma \Delta (\Delta \hat{\theta}_i-\caly_i)\\
\nonumber
\dot w  &= -\gamma \Delta^2 w, \; w(0)=1\\
\hat E &=\xi_E-\Phi {1 \over 1 - w_c}[\hat \theta - w_c \hat \theta(0)],
\lab{ftcstaest}
}
where $w_c$ is defined via the clipping function
$$
w_c = \left\{ \begin{array}{lcl} w & \;\mbox{if}\; & w < 1-\mu \\ 1-\mu & \;\mbox{if}\; & w \geq 1-\mu, \end{array} \right. 
$$
ensures
$$
\tilde E(t)=0,\;\forall t \geq t_c
$$
with all signals bounded, provided  $\Delta(t)$ verifies Assumption \ref{ass2}.
\end{prop}

\begin{pf}
First, notice that the definition of $w_c$ ensures that $\hat E$ is well-defined. Now, from (\ref{tetaSol}) and the definition of $w$ we have that 
$$
\tilde \theta =w\tilde \theta(0).
$$ 
Clearly, this is equivalent to
\begequ
\lab{algrel}
(1 - w)\theta = \hat \theta  - w \hat \theta(0).
\endequ
On the other hand,  under Assumption \ref{ass2}, we have that $w_c(t)=w(t),\; \forall t \geq t_c$. Consequently, we conclude that 
$$
 {1 \over 1 - w_c}[\hat \theta - w_c \hat \theta(0)]=\theta,\; \forall t \geq t_c.
$$
Replacing this identity in \eqref{ftcstaest} completes the proof.
\qed
\end{pf}

The following remarks are in order.

\begin{itemize}
\item[(R1)]  The stability properties established in Propositions \ref{pro1} and \ref{pro2} are ``trajectory-dependent", in the sense that they pertain only to the trajectory generated for the given initial conditions.  This means that the flow of the observer dynamics may contain unbounded trajectories, and the appearance of a perturbation may drive our ``good" trajectory towards a ``bad" one. This is, of course, a robustness problem that needs to be further investigated.

\item[(R2)] It is interesting to note that the ``excitation" condition imposed by Assumption \ref{ass2} can be weakened increasing the adaptation gain $\gamma$. However, as is well-known, the use of fast adaptation entails a series of undesirable phenomena. On the other hand, if the  assumption of  $\Delta \notin \call_2$ is strengthened to $\Delta$ being PE, then the convergence is {\em exponential}.
\end{itemize}
%
%%%%%%%%%%%%

\section{Simulation Results}
\lab{sec4}
%%%%%%%%%%%%%%
%
For simulation we consider a two-machine system, see for instance \citep{ ORTetal}. The dynamics of the system result in the sixth-order model
\begin{footnotesize}
\begin{align}
\label{2masch}
\begin{cases}
\dot{\delta}_1 &=\omega_1, \\
\dot{\omega}_1 &=-D_1 \omega_1 + P_1 - G_{11} E_1^2 - Y_{12} E_1 E_2 \sin(\delta_{12}+\alpha_{12})\\
\dot{E}_1 &= -a_1 E_1 + b_1 E_2 \cos(\delta_{12}+\alpha_{12}) + E_{f_1} + \nu_1;\\
\dot{\delta}_2 &=\omega_2, \\
\dot{\omega}_2 &=-D_2 \omega_2 + P_2 - G_{22} E_2^2 + Y_{21} E_1 E_2 \sin(\delta_{12}+\alpha_{12})\\
\dot{E}_2 &= -a_2 E_2 + b_2 E_1 \cos(\delta_{21}+\alpha_{21}) + E_{f_2} + \nu_2,
\end{cases}
\end{align}
\end{footnotesize}
with the current equations defined as 
\begin{align}
I_{q1}&=G_{11}E_{1}+ E_{2} Y_{12} \sin(\delta_{12}+\alpha_{12})\\
I_{d1}&=-B_{11}E_{1}-E_{2} Y_{12} \cos(\delta_{12}+\alpha_{12})\\
I_{q2}&=G_{22}E_{2}+ E_{1} Y_{21} \sin(\delta_{21}+\alpha_{21})\\
I_{d2}&=-B_{22}E_{2}-E_{1} Y_{21} \cos(\delta_{21}+\alpha_{21}).
\end{align} 
In this case we have that
\begalis{
A(t)&=\begmat{ -a_1 & b_1  \cos(\delta_{12}(t)+\alpha_{12}) \\ b_2  \cos(\delta_{21}(t)+\alpha_{21})& -a_2 }\\
S(\delta)&= \begmat{ G_{11} & Y_{12} \sin(\delta_{12}+\alpha_{12})\\  Y_{21} \sin(\delta_{21}+\alpha_{21}) & G_{22}}\\
T(\delta)&= \begmat{- B_{11} & - Y_{12} \cos(\delta_{12}+\alpha_{12})\\  - Y_{21} \cos(\delta_{21}+\alpha_{21}) & - B_{22}}.
}
For the observer design we selected the simplest filter
$$
F(p)=\begmat{1 & 0 \\ \frac{k}{p+k} & 0},
$$
with $k>0$. The parameters of the model \eqref{2masch} are given in Table \ref{table_1}.

\begin{table}[h]
\caption{System parameters}
\label{table_1}
\begin{center}
\begin{tabular}{|c|c|c|}
\hline
Parameter & Initial values & After load change\\
\hline
$Y_{12}$ & 0.1032 &  0.1032\\
\hline
$Y_{21}$ & 0.1032 & 0.1032\\
\hline
$b_1$ &0.0223 & 0.02236\\
\hline
$b_2$ &0.0265 & 0.0265\\
\hline
$D_1$ &1 & 1\\
\hline
$D_2$ &0.2 & 0.2\\
\hline
$\nu_1$ &1 & 1\\
\hline
$\nu_2$ & 1 & 1\\
\hline
$B_{11}$ & -0.4373 & -0.5685\\
\hline
$B_{22}$ & -0.4294 & -0.5582\\
\hline
$G_{11}$ & 0.0966 & 0.1256 \\
\hline
$G_{22}$ &0.0926 &0.1204 \\
\hline
$a_1$ & 0.2614 & 0.2898\\
\hline
$a_2$ & 0.2532 & 0.2864\\
\hline
$P_1$ & 28.22 & 28.22\\
\hline
$P_2$  & 28.22 & 28.22\\
\hline
$E_{f1}$ & 0.2405 & 0.2405\\
\hline
$E_{f2}$ & 0.2405 & 0.2405\\
\hline
\end{tabular}
\end{center}
\end{table}
 
\begin{figure}
\begin{center}
	\includegraphics[width=9.4cm]{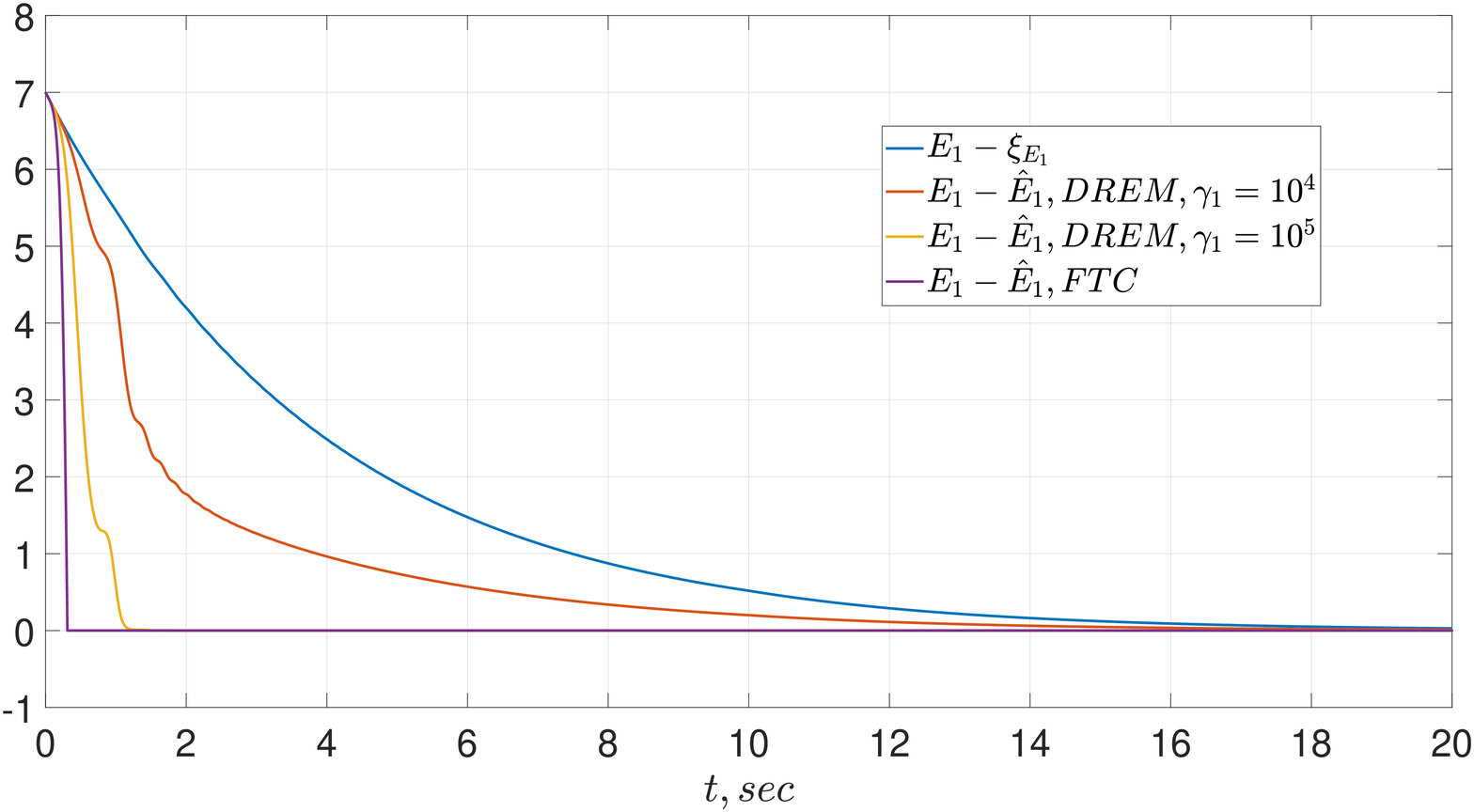}    
	\caption{Transients of the first voltage observation error $E_1-\hat{E}_1$ for DREM and FTC observers with a 30\% load change a $t=10$ sec
	}
	\label{fig:pic2}
\end{center}
\end{figure}

\begin{figure}
\begin{center}
	\includegraphics[width=9.4cm]{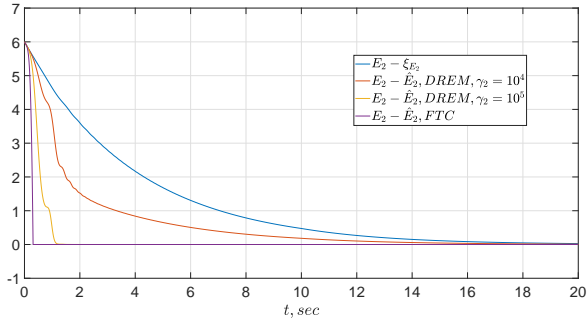}    
	\caption{Transients of the second voltage observation error $E_2-\hat{E}_2$ for DREM and FTC observers with a 30\% load change at $t=10$ sec
	}
	\label{fig:pic4}
\end{center}
\end{figure}

\begin{figure}
\begin{center}
	\includegraphics[width=9.4cm]{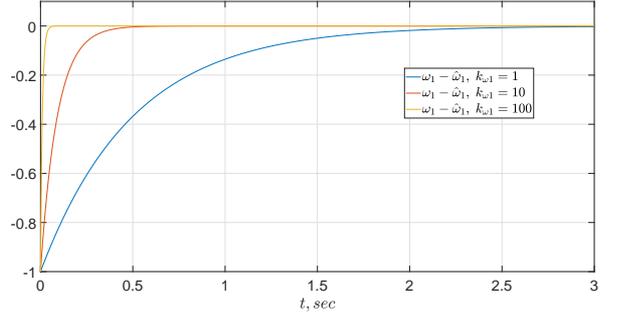}    
	\caption{Transients of the first speed observation error $\omega_1-\hat{\omega}_1$
	}
	\label{fig:pic5}
\end{center}
\end{figure}

\begin{figure}
\begin{center}
	\includegraphics[width=9.4cm]{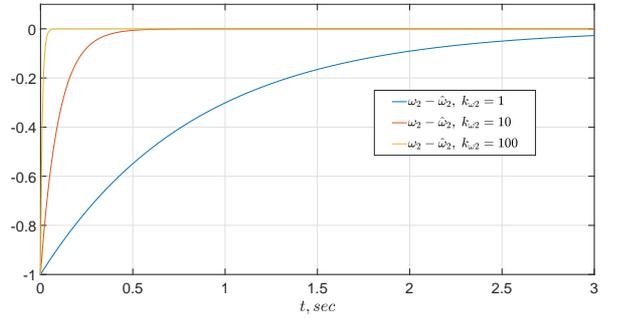}    
	\caption{Transients of the second speed observation error $\omega_2-\hat{\omega}_2$ 
	}
	\label{fig:pic6}
\end{center}
\end{figure}

Simulations were carried out for the observers proposed in Proposition \ref{pro1} (DREM) and  Proposition \ref{pro2} (FTC). We consider the realistic scenario of a  30\% load change, that happens at $t=10$ sec.  It is clear from the proof of Proposition \ref{pro1} that this load change does not affect the speed observation. We used the following initial conditions for the system $E(0)=\col(7,6)$, and zero for all remaining states. All initial conditions for the observers were also set equal to zero. For the DREM observer two different adaptation gains $\gamma_i$ were tried. For the FTC observer we set $\mu=0.1$ of Assumption \ref{ass2}. For the speed observer we selected different values of the coefficient $k_{\omega i}$. For the sake of comparison we also present the behavior of the signals $\xi_{Ei}$ that---since the matrix $A(t)$ corresponds to an asymptotically stable system---provides also an estimate of the voltage $E$. The results of the simulations are shown in Fig.~\ref{fig:pic2}-Fig.~\ref{fig:pic6}. As expected by the theory, the speed of convergence of the DREM estimator for the voltage increases for large values of $\gamma_i$, with a similar behavior of the speed error with respect to $k_{\omega i}$. Also, it is seen that the FTC estimator has the fastest convergence, and this happens in finite-time. It is interesting to remark that the behavior of the observers is highly insensitive to the load change, as it is hardly visible in $\tilde E_2$.

%%%%%%%%%%%%%%%%
\section{Concluding Remarks and Future Research}
\lab{sec5}
%%%%%%%%%%%%%%
%
We have given in this paper a solution to the problem of state observation of multimachine power systems with lossy lines assuming measurable the rotor angles, the excitation voltage and the active and reactive power of all generators. To the best of the authors' knowledge this is the first time a solution like this is presented. It is shown that, under a suitable observability assumption, global convergence of the voltage estimator is achieved. By imposing the classical PE condition, the convergence is exponential ensuring good robustness properties. The speed observation is ensured without imposing any excitation condition, and it is exponential with tunable and, possibly, arbitrarily fast rate of convergence. 

Besides its direct application in (unrealistic) state-feedback excitation controllers for transient stability improvement, {\em e.g.} \citep{ ORTetal,SHEetal,LANetal,YANetal}, the present result may be utilized by the wide-area control system to control the transient and oscillatory dynamics of system voltage and frequency \citep{ singh2018dynamic}. 

Several open topics remain to be investigated.
\begin{itemize}
\item A better understanding of the key excitation assumption $\Delta \notin \call_2$ is required. In particular, to know if it is verified in the normal (or fault) conditions of the power system.
\item As indicated in the introduction a major practical issue is the lack of synchronism in the measurements of the rotor angles. A possible scenario to capture this fact is to include time delays in the transmission information---where the recent results of \citep{ rueda2019gramian} might be useful. 
\item The parameters of the network are highly uncertain. In particular, the line impedances, are poorly known. An adaptive implementation of the proposed observer is currently been explored. 
\end{itemize}
%
%%%%%%%%%%%%%%%%%
\section*{Acknowledgment}
The authors would like to thank  Stanislav Aranovskiy and Denis Efimov for pointing out the LTI formulation of the error dynamics discussed in Subsection \ref{subsec31}. The second author is also grateful to Gerardo Espinosa for bringing to his attention the problem of state estimation of power systems. 
\bibliography{bib_power_observer}

\begin{thebibliography}{38}
\providecommand{\natexlab}[1]{#1}
\providecommand{\url}[1]{\texttt{#1}}
\providecommand{\urlprefix}{URL }
\expandafter\ifx\csname urlstyle\endcsname\relax
  \providecommand{\doi}[1]{doi:\discretionary{}{}{}#1}\else
  \providecommand{\doi}{doi:\discretionary{}{}{}\begingroup
  \urlstyle{rm}\Url}\fi

\bibitem[{Anagnostou et~al.(2018)Anagnostou, Boem, Kuenzel, Pal, and
  Parisini}]{anagnostou18}
Anagnostou, G., Boem, F., Kuenzel, S., Pal, B.C., and Parisini, T. (2018).
\newblock Observer-based anomaly detection of synchronous generators for power
  systems monitoring.
\newblock \emph{IEEE Transactions on Power Systems}, 33(4), 4228--4237.

\bibitem[{Anagnostou and Pal(2017)}]{anagnostou17}
Anagnostou, G. and Pal, B.C. (2017).
\newblock Derivative-free kalman filtering based approaches to dynamic state
  estimation for power systems with unknown inputs.
\newblock \emph{IEEE Transactions on Power Systems}, 33(1), 116--130.

\bibitem[{Anderson(1971)}]{AND}
Anderson, B. (1971).
\newblock Stability properties of kalman-bucy filters.
\newblock \emph{Journal of the Franklin Institute}, 291(2), 137--144.

\bibitem[{Anderson and Fouad(2003)}]{ANDFOU}
Anderson, P.M. and Fouad, A.A. (2003).
\newblock \emph{Observer Design for Nonlinear Systems}.
\newblock Wiley-IEEE Pres.

\bibitem[{Aranovskiy et~al.(2017)Aranovskiy, Bobtsov, and Pyrkin}]{ARAetal}
Aranovskiy, S., Bobtsov, A., and Pyrkin, A. (2017).
\newblock Performance enhancement of parameter estimators via dynamic regressor
  extension and mixing.
\newblock \emph{\TAC}, 62, 3546--3550.

\bibitem[{Bernard(2019)}]{BER}
Bernard, P. (2019).
\newblock \emph{Power System Control and Stability}.
\newblock Springer-Verlag, Vol. 479.

\bibitem[{Besan{\c{c}}on(2007)}]{BES}
Besan{\c{c}}on, G. (2007).
\newblock \emph{Nonlinear observers and applications}, volume 363.
\newblock Springer.

\bibitem[{Demidovich(1967)}]{DEM}
Demidovich, B.P. (1967).
\newblock \emph{Lectures on Mathematical Stability Theory (in Russian)}.
\newblock Nauka.

\bibitem[{Gerasimov et~al.(2018)Gerasimov, Ortega, and Nikiforov}]{GERetal}
Gerasimov, D., Ortega, R., and Nikiforov, V. (2018).
\newblock Adaptive control of multivariable systems with reduced knowledge of
  high frequency gain: Application of dynamic regressor extension and mixing
  estimators.
\newblock \emph{IFAC-PapersOnLine}, 5(1), 886--890.

\bibitem[{Ghahremani and Kamwa(2011)}]{ghahremani11}
Ghahremani, E. and Kamwa, I. (2011).
\newblock Dynamic state estimation in power system by applying the extended
  kalman filter with unknown inputs to phasor measurements.
\newblock \emph{IEEE Transactions on Power Systems}, 26(4), 2556--2566.

\bibitem[{Julier and Uhlmann(1997)}]{julier97}
Julier, S.J. and Uhlmann, J.K. (1997).
\newblock New extension of the {Kalman} filter to nonlinear systems.
\newblock In \emph{Signal processing, sensor fusion, and target recognition
  VI}, volume 3068, 182--193. International Society for Optics and Photonics.

\bibitem[{Kreisselmeier and Rietze-Augst(1990)}]{KRERIE}
Kreisselmeier, G. and Rietze-Augst, G. (1990).
\newblock Richness and excitation on an interval---with application to
  continuous-time adaptive control.
\newblock \emph{\TAC}, 35(2), 165--171.

\bibitem[{Kundur(1994)}]{kundur94}
Kundur, P. (1994).
\newblock \emph{Power System Stability and Control}.
\newblock McGraw-Hill.

\bibitem[{Langarica et~al.(2015)Langarica, Ortega, and Casagrande}]{LANetal}
Langarica, D., Ortega, R., and Casagrande, D. (2015).
\newblock Transient stability of multimachine power systems: towards a global
  decentralized solution.
\newblock \emph{European Journal of Control}, 26, 44--52.

\bibitem[{Ljung(1987)}]{LJU}
Ljung, L. (1987).
\newblock \emph{System Identification: Theory for the User}.
\newblock Prentice Hall, New Jersey.

\bibitem[{Milano et~al.(2018)Milano, D{\"o}rfler, Hug, Hill, and
  Verbi\v{c}}]{milano18}
Milano, F., D{\"o}rfler, F., Hug, G., Hill, D.J., and Verbi\v{c}, G. (2018).
\newblock Foundations and challenges of low-inertia systems.
\newblock In \emph{2018 Power Systems Computation Conference (PSCC)}, 1--25.
  IEEE.

\bibitem[{Ortega et~al.(2019)Ortega, Bobtsov, Dochain, and
  Nikolaev}]{ORTetaljpc}
Ortega, R., Bobtsov, A., Dochain, D., and Nikolaev, N. (2019).
\newblock State observers for reaction systems with improved convergence rate.
\newblock \emph{\JPC}, 83, 53--62.

\bibitem[{Ortega et~al.(2015)Ortega, Bobtsov, Pyrkin, and
  Aranovskyi}]{ORTetalscl}
Ortega, R., Bobtsov, A., Pyrkin, A., and Aranovskyi, S. (2015).
\newblock A parameter estimation approach to state observation of nonlinear
  systems.
\newblock \emph{Systems and Control Letters}, 85, 84--94.

\bibitem[{Ortega et~al.(2005)Ortega, Galaz, Astolfi, Sun, and Shen}]{ORTetal}
Ortega, R., Galaz, M., Astolfi, A., Sun, Y.Z., and Shen, T. (2005).
\newblock Transient stabilization of multimachine power systems with nontrivial
  transfer conductances.
\newblock \emph{\TAC}, 50(1), 1--16.

\bibitem[{Paul et~al.(2018)Paul, Joos, and Kamwa}]{paul18}
Paul, A., Joos, G., and Kamwa, I. (2018).
\newblock Dynamic state estimation of full power plant model from terminal
  phasor measurements.
\newblock In \emph{2018 IEEE/PES Transmission and Distribution Conference and
  Exposition (T\&D)}, 1--5. IEEE.

\bibitem[{Pyrkin et~al.(2019)Pyrkin, Bobtsov, Ortega, Vedyakov, and
  Aranovskiy}]{PYRetal}
Pyrkin, A., Bobtsov, A., Ortega, R., Vedyakov, A., and Aranovskiy, S. (2019).
\newblock Adaptive state observer design using dynamic regressor extension and
  mixing.
\newblock \emph{\SCL}, 133, pp. 1--8.

\bibitem[{Rinaldi et~al.(2017)Rinaldi, Cucuzzella, and Ferrara}]{rinaldi17}
Rinaldi, G., Cucuzzella, M., and Ferrara, A. (2017).
\newblock Third order sliding mode observer-based approach for distributed
  optimal load frequency control.
\newblock \emph{IEEE Control Systems Letters}, 1(2), 215--220.

\bibitem[{Rinaldi et~al.(2018)Rinaldi, Cucuzzella, and Ferrara}]{rinaldi18}
Rinaldi, G., Cucuzzella, M., and Ferrara, A. (2018).
\newblock Sliding mode observers for a network of thermal and hydroelectric
  power plants.
\newblock \emph{Automatica}, 98, 51--57.

\bibitem[{Rinaldi et~al.(2019)Rinaldi, Menon, Edwards, and Ferrara}]{rinaldi19}
Rinaldi, G., Menon, P.P., Edwards, C., and Ferrara, A. (2019).
\newblock Higher order sliding mode observers in power grids with traditional
  and renewable sources.
\newblock \emph{IEEE Control Systems Letters}, 4(1), 223--228.

\bibitem[{Rueda-Escobedo et~al.(2019)Rueda-Escobedo, Ushirobira, Efimov, and
  Moreno}]{rueda2019gramian}
Rueda-Escobedo, J.G., Ushirobira, R., Efimov, D., and Moreno, J.A. (2019).
\newblock Gramian-based uniform convergent observer for stable ltv systems with
  delayed measurements.
\newblock \emph{International Journal of Control}, 1--12.

\bibitem[{Rugh(1996)}]{RUG}
Rugh, W. (1996).
\newblock \emph{Linear System Theory}.
\newblock Prentice Hall, New Jersey, USA,.

\bibitem[{Sastry and Bodson(1989)}]{SASBOD}
Sastry, S. and Bodson, M. (1989).
\newblock \emph{Adaptive Control: Stability, Convergence and Robustness}.
\newblock Prentice Hall, New Jersey, USA.

\bibitem[{Sauer et~al.(2017)Sauer, Pai, and Chow}]{sauer17}
Sauer, P.W., Pai, M.A., and Chow, J.H. (2017).
\newblock \emph{Power system dynamics and stability: with synchrophasor
  measurement and power system toolbox}.
\newblock John Wiley \& Sons.

\bibitem[{Shen et~al.(2003)Shen, Ortega, Lu, Mei, and Tamura}]{SHEetal}
Shen, T., Ortega, R., Lu, Q., Mei, S., and Tamura, K. (2003).
\newblock Adaptive disturbance attenuation of hamiltonian systems with
  parametric perturbations and application to power systems.
\newblock \emph{Asian J of Control}, 5(1).

\bibitem[{Singh and Pal(2018)}]{singh2018dynamic}
Singh, A.K. and Pal, B. (2018).
\newblock \emph{Dynamic Estimation and Control of Power Systems}.
\newblock Academic Press.

\bibitem[{Terzija et~al.(2010)Terzija, Valverde, Cai, Regulski, Madani, Fitch,
  Skok, Begovic, and Phadke}]{terzija10}
Terzija, V., Valverde, G., Cai, D., Regulski, P., Madani, V., Fitch, J., Skok,
  S., Begovic, M.M., and Phadke, A. (2010).
\newblock Wide-area monitoring, protection, and control of future electric
  power networks.
\newblock \emph{Proceedings of the IEEE}, 99(1), 80--93.

\bibitem[{Valverde and Terzija(2011)}]{valverde11}
Valverde, G. and Terzija, V. (2011).
\newblock Unscented {Kalman} filter for power system dynamic state estimation.
\newblock \emph{IET generation, transmission \& distribution}, 5(1), 29--37.

\bibitem[{Wan and Van Der~Merwe(2000)}]{wan00}
Wan, E.A. and Van Der~Merwe, R. (2000).
\newblock The unscented {Kalman} filter for nonlinear estimation.
\newblock In \emph{Proceedings of the IEEE 2000 Adaptive Systems for Signal
  Processing, Communications, and Control Symposium (Cat. No. 00EX373)},
  153--158.

\bibitem[{Wang et~al.(2011)Wang, Gao, and Meliopoulos}]{wang11}
Wang, S., Gao, W., and Meliopoulos, A.S. (2011).
\newblock An alternative method for power system dynamic state estimation based
  on unscented transform.
\newblock \emph{IEEE transactions on power systems}, 27(2), 942--950.

\bibitem[{Winter et~al.(2015)Winter, Elkington, Bareux, and Kostevc}]{winter15}
Winter, W., Elkington, K., Bareux, G., and Kostevc, J. (2015).
\newblock Pushing the limits: Europe's new grid: Innovative tools to combat
  transmission bottlenecks and reduced inertia.
\newblock \emph{IEEE Power and Energy Magazine}, 13(1), 60--74.

\bibitem[{Yan et~al.(2010)Yan, Dong, Saha, and Majumder}]{YANetal}
Yan, R., Dong, Z., Saha, T., and Majumder, R. (2010).
\newblock A power system nonlinear adaptive decentralized controller design.
\newblock \emph{Automatica}, 46, 330--336.

\bibitem[{Yang et~al.(2007)Yang, Bi, and Wu}]{yang07}
Yang, Q., Bi, T., and Wu, J. (2007).
\newblock Wams implementation in china and the challenges for bulk power system
  protection.
\newblock In \emph{2007 IEEE Power Engineering Society General Meeting}, 1--6.
  IEEE.

\bibitem[{Zhao et~al.(2019)Zhao, Gomez-Exposito, Netto, Mili, Abur, Terzija,
  Kamwa, Pal, Singh, Qi et~al.}]{zhao19}
Zhao, J., Gomez-Exposito, A., Netto, M., Mili, L., Abur, A., Terzija, V.,
  Kamwa, I., Pal, B.C., Singh, A.K., Qi, J., et~al. (2019).
\newblock Power system dynamic state estimation: motivations, definitions,
  methodologies and future work.
\newblock \emph{IEEE Transactions on Power Systems}.

\end{thebibliography}
\end{document}